\documentclass[twocolumn,showpacs,preprintnumbers,amsmath,amssymb]{revtex4}

\usepackage{graphicx}
\usepackage{dcolumn}
\usepackage{bm}

\begin{document}

\title{Transitions To the Long-Resident State in coupled chaotic oscillators}
\author{Ao Bin and Zheng Zhi-Gang\footnote{Corresponding author: zgzheng@bnu.edu.cn}}
\address{
Department of Physics and the Beijing-Hong-Kong-Singapore Joint
Center for Nonlinear and Complex Systems (Beijing), Beijing Normal
University, Beijing 100875, CHINA }

\begin{abstract}
The behaviors of coupled chaotic oscillators before complete
synchronization were investigated. We report three phenomena: (1)
The emergence of long-time residence of trajectories besides one
of the saddle foci; (2) The tendency that orbits of the two
oscillators get close becomes faster with increasing the coupling
strength; (3) The diffusion of two oscillator's phase difference
is first enhanced and then suppressed. There are exact
correspondences among these phenomena. The mechanism of these
correspondences is explored. These phenomena uncover the route to
synchronization of coupled chaotic oscillators.
\end{abstract}

\maketitle

Synchronization is a universal and fundamental behavior occuring
in various fields\cite{bsyn,bzh,xu2004}. Different synchronous
states can be observed in coupled chaotic systems such as complete
synchronization (CS)\cite{4,shi},generalized
synchronization(GS)\cite{gs_zhang}, phase synchronization
(PS)\cite{5,qian}, measure synchronization (MS)\cite{ms_chen} and
so on. What people care about in studies of synchronization is
often the critical coupling, in which the synchronization can
achieve\cite{6}, and the behavior near this threshold. The dynamic
in regimes far from the global synchronous state is related to the
mechanism that makes the synchronization be a stable state. In
this paper we report three phenomena which can help us to
understand the problem above. We also uncover a mechanism of
synchronization in systems which have two or more saddle foci.

When two or more chaotic systems were linearly coupled into a
network, their differential equation can be represented by
$\dot{\vec{X}}=\vec{F}(\vec{X})+\varepsilon\Gamma\otimes C
\vec{X}$, where $\vec{X}=(\vec{x}^1,\vec{x}^2,\cdots,\vec{x}^N)$,
and $N$ is the node number of network. $\varepsilon$ denotes the
coupling strength. $\Gamma:\mathbb{R}^m \rightarrow \mathbb{R}^m$
characterizes the coupling schemes among the variables of the
nodes in a network. $C=M-D$, where $M$ denotes the adjacency
matrix of the network (the element $M_{ij}$ denotes the number of
the edges that link node $i$ and $j$), $D$ is a diagonal matrix,
and satisfies $D_{ii}=\sum_{j=1}^{N}M_{ij}$. In this paper, we
take the $N$ Lorenz systems ($\dot{x}=\sigma(y-x),\dot{y}=rx-y-xz,
\dot{z}=xy-\beta z,$ where $\sigma=10, r=28, \beta=8/3$) as our
nodes and couple them into an array (the periodic boundary
condition is applied) with $\Gamma_{ij}=0$ $(i,j=1,2,3)$ except
$\Gamma_{11}=1$.

\begin{figure}
\includegraphics[width=3.3in,height=2.6in]{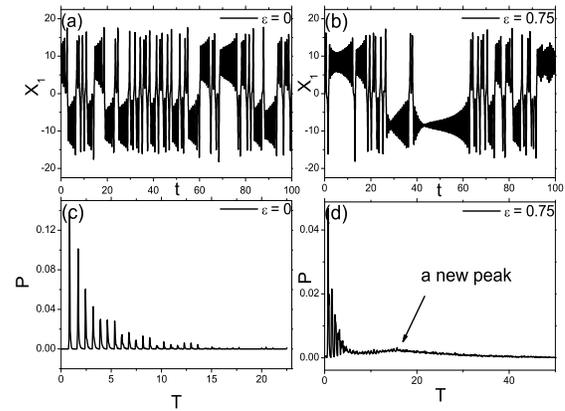}
\caption{\label{f1}(a):Time series $x_i(t)$ for a single Lorenz
system. (b):Residence time statistics $P(T)$. (c)(d):Coupled
Lorenz system($N=2$),$\varepsilon=0.75$.}
\end{figure}

\begin{figure}
\includegraphics[width=3.3in,height=2.4in]{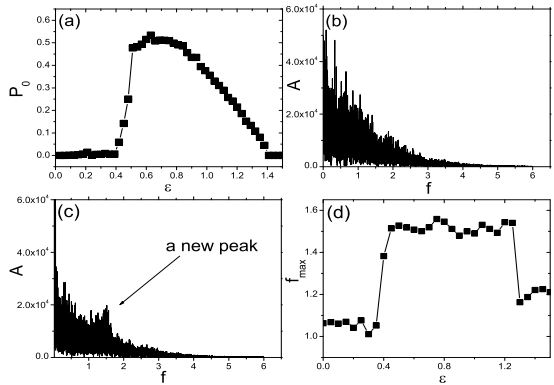}
\caption{\label{f2}(a):$P_0 \sim \varepsilon,$ $T_0=10$. (b):Power
spectrum $A(f)$ of $x(t)$ for a single Lorenz system. (c):$A(f)$
of $x_1(t)$ for coupled Lorenz system($N=2$), $\varepsilon=0.75$.
(d):$f_{max} \sim \varepsilon, f_0=1$.}
\end{figure}

There are three fixed points in phase space of a single Lorenz
oscillator. Two of them are saddle foci
($\pm\sqrt{\beta(r-1)}$,$\pm\sqrt{\beta(r-1)}$,$r-1$) and another
is a saddle node(0,0,0). The trajectory usually revolves around
the two saddle foci alternatively. Fig.\ref{f1}(a) is the time
serial of $x$-component of a single Lorenz system. The revolution
about one of the saddle foci can be regarded as one dynamic phase.
Fig.\ref{f1}(b) is the statistics $P(T)$ of the residence time $T$
that the system resides in a phase. This distribution consists of
several almost discrete peaks. When the two systems are coupled
together, new phenomenon can be observed. Fig.\ref{f1}(c) is one
of the time series of $x$-component in a system consisting of two
Lorenz oscillators. It can be observed that the residence time in
certain phase is longer than the situation in Fig.\ref{f1}(a).
Accordingly, a new wide continuous smooth peak appears in the
distribution of residence time in Fig.\ref{f1}(d). The above
phenomenon suggests a new state that we call long-residence
state(LRS) in coupled chaotic systems. While the new peak in the
distribution indicate the emergence of LRS, one can define
$P_0=\int_{T_0}^{+\infty}P(T)dT$ as the probability of resident
time that is larger than $T_0$. For a larger $T_0$, $P_0$ is a
measure of the emergence of LRS. Fig.\ref{f2}(a) is the
relationship between $\varepsilon$ and $P_0$($T_0=10$). Obviously,
for a weak coupling, $P_0\approx0$; But when the coupling
$\varepsilon$ is greater than a critical point
$\varepsilon_1\approx0.36$, $P_0$ increase sharply, suggesting the
emergence of the LRS.

On the other hand, the emergence of LRS can also be observed
through the power spectrum of the chaotic motion. Fig.\ref{f2}(b)
is the spectrum amplitude of a single Lorenz system, and
Fig.\ref{f2}(c) gives that of a coupled system with two
oscillators. A new peak appears in the high-frequency regime. The
relationship between this frequency peak in the region
$[f_0,\infty)$ and coupling strength $\varepsilon$ is plotted in
Fig.\ref{f2}(d). A critical coupling strength $\varepsilon_1$ can
also be found, indicating the emergence of LRS, and the critical
point is same as that in Fig.\ref{f2}(a).

\begin{figure}
\includegraphics[width=3.5in,height=2in]{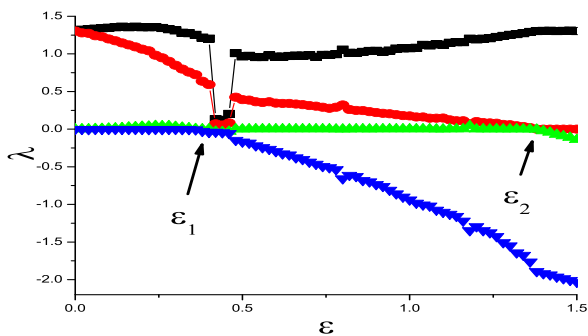}
\caption{\label{f3}The four largest LEs for $N=2$ Lorenz
oscillators. There are two critical coupling strength
$\varepsilon_1=0.36$ and $\varepsilon_2=1.38$.}
\end{figure}

\begin{figure}
\includegraphics[width=3.5in,height=2.2in]{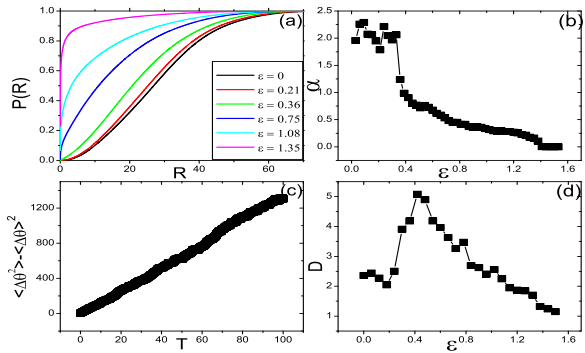}
\caption{\label{f4} $N=2$ (a):$P(R) \sim R$.
(b):$\alpha\sim\varepsilon.$ (c):The behavior of
$\langle\Delta\theta^2\rangle-\langle\Delta\theta\rangle^2$. (d)
The diffusion coefficient $D$ against the coupling
strength$\varepsilon$.}
\end{figure}

To uncover what happens near the critical point of LRS, it is
instructive to study the Lyapunov exponent(LE) spectrum is useful.
Fig.\ref{f3} is the four largest LEs of two coupled Lorenz
oscillators. At $\varepsilon_1$=0.36 and $\varepsilon_2$=1.38, one
of the zero exponents is found to become negative respectively.
Therefore, the LRS is accompanied by a topological transition of
the chaotic attractor and is an intrinsic bifurcation embedded in
complicated motion. One of the Lyapunov exponents becoming
negative means the decrease of the dimension of the chaotic
attractor. On the other hand, when CS achieve at $\varepsilon_2$,
the state of the coupled system is on an invariant synchronous
sub-manifold, and the dimension of the attractor will become one
half of that for uncoupled system. Are there any relations between
LRS and the change of the Lyapunov exponents. To answer the
question, let's study the evolution of the trajectory distance
between two oscillators. We define the Euclidean distance as
$r(t)=\sqrt{(x_2-x_1)^2+(y_2-y_1)^2+(z_2-z_1)^2}$ . When the
synchronization is achieved, $r(t)$ will tend to 0 rapidly. In the
non-synchronous range, the behavior of $r(t)$ is complex, and we
prefer to studying its statistical characters. By defining
$p(r)dr$ as the probability of the distance $r$ located in $r
\rightarrow r + dr$, the accumulative distribution
$P(R)=\int_0^Rp(r)dr$ describes the portion of trajectory distance
that is smaller than R. In the non-synchronous regime
$P(R)\rightarrow 0$ for $R\rightarrow 0$, and $P(R)\rightarrow 1$
when $R\rightarrow R_{max}$, where $R_{max}$ is the determined by
the attractor size. When CS is achieved, one has $P(R) = 1$.
Fig.\ref{f4}(a) shows the behavior of $P(R)$ for different
coupling strengths. It is interesting that $P(R)$ obeys a power
law for $R \ll 1$, i.e. $P(R)\propto R^\alpha$. This indicates
$p(r)\propto r^{\alpha-1}$ , implying $p(r)$ may change from a
Gaussian-like function to a Poisson-like function when $\alpha$
decreases below 1($p(r) \rightarrow 0$ when $r \rightarrow
R_{max}$). The variation of the exponent versus the coupling
$\varepsilon$ is shown in Fig.\ref{f4}(b). It is very interesting
to notice that at $\varepsilon<\varepsilon_1$, $\alpha>1$, and
when $\varepsilon>\varepsilon_1$, $\alpha<1$. This manifests the
emergence of the long-resident state.

the emergence of LRS implies longer rotations before switching to
another rotating saddle focus. This should be closely related to
the phase dynamics of coupled Lorenz oscillators. In Lorenz
system, a dynamical phase $\theta_i=tg^{-1}[u_y^i(t)/u_x^i(t)]$
can be introduced, where $u_x^i(t)=z_i-(r-1)$ and
$u_y^i(t)=\sqrt{x^2_i+y^2_i}-\sqrt{2\beta(r-1)},$$(i=1,2,\cdots,N)$.
The definition of PS is that the average frequency of two or more
oscillator is equal to each other, so the synchronization is in
terms of the "average" of the phase. In fact, the phase defined
above is not a well-defined quantity(no better definition of a
single phase been proposed for oscillations with multiple rotation
centers). Therefore diffusion process can be observed for the
phase difference between tow oscillators,
i.e.$\Delta\theta=\theta_i-\theta_j$,$i,j=1,2,\cdots,N$ are
time-dependent. To understand the changes of diffusive process, we
calculate the second center moment of the difference
$\langle\Delta\theta^2\rangle-\langle\Delta\theta\rangle^2$, where
$\langle\cdot\rangle$ means the average of ensembles.
Fig.\ref{f4}(c) describes the evolution of the difference at
$\varepsilon=0.75$. It is approximately proportional to time,
therefore this process is the Brownian motion. We can further
define the diffusion coefficient,
$$D=\lim\limits_{T\to\infty}\frac{1}{2T}[\langle\Delta\theta^2\rangle-\langle\Delta\theta\rangle^2]$$
In Fig.\ref{f4}(d), we give the behavior of the diffusion
coefficient $D$ against the coupling $\varepsilon$. It can be
found that the diffusion is first enhanced for weak couplings and
then depressed for larger couplings. There is a peak near the
coupling $\varepsilon_1$ where the LRS emerges. This
correspondence implies that the appearance of LRS does favor to
PS. Because there are two rotating centers in a single Lorenz
system, a small coupling will do harm to the synchronization of
coupled system\cite{8}. When the system is in the LRS, oscillators
usually prefer to staying in one of the phase, and the
trajectories of two oscillators are easy to close.

\begin{figure}
\includegraphics[width=3.5in,height=2.5in]{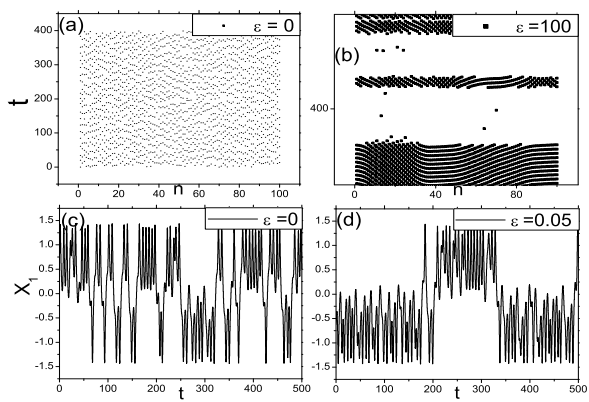}
\caption{\label{f5}(a)(b):The spatiotemporal behavior of a lattice
of $N=100$, $\varepsilon=0$[(a)], $\varepsilon=100$[(b)]. (c)(d):
The evolution of $x(t)$ for two coupled Duffing oscillators(
$\ddot{x}+0.5\dot{x}+x^3-x-0.4cos(t)=0$) with different
$\varepsilon$.}
\end{figure}

In fact, a system with more oscillators may exhibits stronger
effect of LRS. Figs.\ref{f5}(a),(b) gives the spatiotemporal
patterns at $\varepsilon=0$ and $\varepsilon=100$ respectively,
for $N=100$. The oscillator is labeled when it is in one scroll.
One can find that the spatiotemporal pattern at $\varepsilon=0$ is
a random pattern, but obviously a very long stay of oscillators in
one scroll can be found for $\varepsilon=100$.

Finally, we want to stress that the LRS we observed here is
generic. It can be observed in different situations. We have
checked coupled Duffing oscillators(as shown in Figs.\ref{f5}(c)
and (d), the residence time of two coupled Duffing oscillators is
obviously longer than that of a single oscillator), Chua circuits,
and Chen's oscillators \cite{bfo}. All these systems exhibit the
LRS behaviors for moderate couplings. Even if there are some
parameter mismatches, a LRS can still be observed. Furthermore, we
also find that the LRS does not depend on the coupling forms,
e.g., $x-, y-$ or $z-$coupling, and local or global.

In conclusion, we explored the dynamics of coupled chaotic
oscillators with multiple scrolls(saddle foci) prior to global
synchronization. We find a generic transition to the long-resident
state (LRS), where the oscillations experience a rather long
duration in one scroll. This transition is manifested by several
different behaviors. First, one zero LE in LE spectrum become
negative. Second, the transition to LRS is accompanied by the
qualitative changes in the trajectory-distance distribution and
power spectra. Third, we show that the emergence of the LRS is
accompanied by the enhancement-depression transition of the
diffusion of the phase difference. The phenomenon of LRS is
generic, i.e., it is irrelevant to the dynamics, number, and the
interaction types of oscillators.


\end{document}